\documentstyle[preprint,aps,epsfig,psfig]{revtex}
\begin{document}

\draft  

\title{\bf On the Mixing of 
the Scalar Mesons $f_0(1370)$, $f_0(1500)$ and $f_0(1710)$}
\vspace{1cm}
\author{\small De-Min Li$^1$\footnote{ E-mail: 
lidm@hptc5.ihep.ac.cn},~ Hong Yu$^{1,2,3}$~ and~ Qi-Xing Shen$^{1,2,3}$ \\ 
\small $^1$Institute of High Energy Physics, Chinese Academy of 
Sciences, \\
\small P.O.Box 918~(4), Beijing $100039$, China\footnote{Mailing 
address}\\ 
\small $^2$CCAST (World Lab), P.O.Box 8730, Beijing $100080$, China\\
\small $^3$Institute of Theoretical Physics, Chinese Academy of Sciences,
Beijing $100080$, China}
\date{\today}
\maketitle
\vspace*{0.3cm}

\begin{abstract}

Based on a $3\times3$ mass matrix describing the mixing of the scalar states 
$f_0(1370)$, $f_0(1500)$ and $f_0(1710)$, the hadronic decays of the 
three states are investigated. Taking into account the two possible 
assumptions concerning the mass 
level order of the bare states 
$|N\rangle=|u\bar{u}+d\bar{d}\rangle/\sqrt{2}$, 
$|S\rangle=|s\bar{s}\rangle$ and $|G\rangle=|gg\rangle$ in the scalar 
sector, 
$M_G > M_S > M_N$ and $M_G > M_N > M_S$, we obtain the glueball-quarkonia 
content 
of the three states by solving the unlinear equations. Some 
predictions about the decays of the three states in two cases are 
presented, which can provide a stringent consistency 
check of the two assumptions. 

\end{abstract}

\vspace{0.2cm}



\newpage

\tighten

\section*{I. Introduction}
\indent

The existence of glueball states made of gluons is 
one of 
the important predictions of QCD. Discovery and confirmation of 
these glueball states would
be the strong support to the QCD theory. Therefore, the 
search for and identifying the glueball states have been a very excited 
and attractive research subject.
The abundance of $q\bar{q}$ mesons and the 
possible mixing of glueballs and ordinary mesons make  
the current situation with the identification of the glueball states  
rather complicated. However, some progress has been made in the glueball 
sector. By studying the mixing between quarkonia and glueball 
to study the properties of glueballs or 
identify the glueball states is an appealing approach 
\cite{Schn,Rosner,Kawai,Tuan,Caruso}.  

In contrast to the vector and tensor mesons, the identification of the 
scalar mesons is a long standing puzzle. In particular, the $I=0$, 
$J^{PC}=0^{++}$ sector is the most complex one both 
experimentally and theoretically. The quark model predicts that there are 
two mesons with $I=0$ in the ground $^3P_0$ $q\bar{q}$ nonet, but apart from 
the state 
$f_J(1710)$ with $J=0$ or ( and ) $2$, four states $f_0(400-1200)$, 
$f_0(980)$, $f_0(1370)$ and $f_0(1500)$ are listed by Particle Data Group 
(PDG)\cite{PDG}. There are too many 
controversies about these states, especially relating
to the $f_0(400-1200)$ and $f_0(980)$. The convenient 
but not convincing recent tendency is to put aside the $f_0(400-1200)$
and $f_0(980)$, and focus on $f_0(1370)$, $f_0(1500)$ and $f_0(1710)$ 
\cite{Seith}, although the spin-parity of the $f_J(1710)$ $J^{PC}=0^{++}$ 
or ( and ) $2^{++}$ is controversial.  

Recently, several authors have discussed the quarkonia-glueball 
content of the $f_0(1370)$, $f_0(1500)$ and $f_0(1710)$ by studying the 
mixing of the three states \cite{CLOSE,Wein,Stro,Bura}. The 
different assumption 
about the mass level order of the bare states $|G\rangle=|gg\rangle$, 
$|S\rangle=|s\bar{s}\rangle$ and 
$|N\rangle=|u\bar{u}+d\bar{d}\rangle/\sqrt{2}$ leads to the different 
quantitative predictions about the glueball-quarkonia content of the 
three states. In Ref. \cite{Wein,Stro,Bura} $M_G > 
M_S > M_N$ is assumed and in Ref. \cite{CLOSE} $M_S > M_G > M_N$ is assumed.
The assumption, $M_G>M_{q\bar{q}}$, is consistent with the prediction 
given by lattice 
QCD \cite{Lee} that the bare glueball state has a higher mass than the 
bare quarkonia states. However, without the confirmation that which 
state, $a_0(980)$ or $a_0(1450)$, is the isovector member in 
the ground $^3P_0$ nonet, the level order of the $M_S$ and $M_N$ in 
the scalar sector perhaps still remains open. On one hand, if  
the $a_0(1450)$ is assigned as the isovector member in the ground $^3P_0$ 
nonet, since the $a_0(1450)$ with mass $1474\pm19$ MeV\cite{PDG} 
has a higher mass than the observed isodoublet scalar states
$K^{\ast}_0(1430)$
with mass $1429\pm6$ MeV\cite{PDG}, according to the Gell-Mann- Okubo mass 
formula\cite{Okubo} one would expect 
$M_N>M_S$. On the other hand, if the $a_0(980)$ is assigned as the 
isovector member, one would expect $M_S>M_N$, which is also consistent 
with that the strange quark $s$ has a higher mass than the non-strange quark 
$u$ or $d$ in a constituent quark picture. We believe that 
neither $M_S>M_N$ nor $M_N>M_S$ seems to 
be ruled out in the scalar sector in the current situation.

In this letter, based on the mixing scheme of the $f_0(1370)$, $f_0(1500)$ 
and $f_0(1710)$, we shall discuss the glueball-quarkonia content of the 
three states taking into account the two possible assumptions 
$M_G > M_S > M_N$ and $M_G >M_N > M_S$\footnote{ Assuming $M_S=M_N$ and 
taking $M_1=1.712$ GeV, $M_2=1.5$ GeV, 
when $M_3$ changes from 1.2 to 1.5 
GeV, from Eqs. (2)$\sim$(4), we find that $M_G>0$ requires 
$M_S=M_N=1.5$ 
GeV, which leads to $x_2=y_2=z_2=0$ in Eqs. (5), (6). Therefore, the 
possibility of $M_S=M_N$ can be ruled out.}.  
This paper is organized as follows: In Sect. II, the two-body hadronic 
decays of $f_0(1710)$, $f_0(1500)$ and $f_0(1370)$ are investigated in 
the quarkonia-glueball mixing framework.
The results for two cases $M_G > M_S > M_N$ and $M_G >M_N > M_S$ are 
presented in Sect. III. Our conclusions are reached in Sect. IV.  

\section*{II. Mixing 
scheme of quarkonia and glueball }
 \indent

In the $|G\rangle=|gg\rangle$, $|S\rangle=|s\bar{s}\rangle$, 
$|N\rangle=|u\bar{u}+d\bar{d}\rangle/\sqrt{2}$ basis,
the mass matrix describing the mixing of a glueball and 
quarkonia can be written as follows \cite{Wein}:
\begin{equation}
M=\left( \begin{array}{ccc}
M_G & f & 
\sqrt{2}f\\
f & M_S & 0\\
\sqrt{2}f & 0 & M_N
\end{array}\right),
\end{equation}
where $f=\langle G|M|S\rangle=\langle G|M|N\rangle/\sqrt{2}$ represents the 
flavor independent mixing strength between the glueball and quarkonia 
states. The vanishing off-diagonal elements indicate that there is no 
direct quarkonia mixing which is assumed to be a higher order effect. $M_G$, 
$M_S$ and $M_N$ represent the masses of 
the bare states $|G\rangle$, $|S\rangle$ and $|N\rangle$, respectively.  
Here we assume that the physical states $|f_0(1710)\rangle$, 
$|f_0(1500)\rangle$ and 
$|f_0(1370)\rangle$ are the eigenstates of $M$ with the eigenvalues of  
$M_{1}$, $M_{2}$ and $M_{3}$, respectively ($M_1$, $M_2$ and $M_3$ 
denote the masses of $f_0(1710)$, $f_0(1500)$ and $f_0(1370)$, 
respectively). If 
one defines a $3\times3$ unitary matrix $U$ which transforms the states 
$|G\rangle$, $|S\rangle$ and $|N\rangle$ into the physical states 
$|f_0(1710)\rangle$, 
$|f_0(1500)\rangle$ and $|f_0(1370)\rangle$, then $UMU^{-1}$ must be the 
diagonal matrix with the diagonal elements $M_{1}$,
$M_{2}$ and $M_{3}$, from which one can get the following equations:

\begin{eqnarray}
&~&M_1+M_2+M_3=M_G+M_S+M_N,\\
&~&M_1M_2+M_1M_3+M_2M_3=M_GM_S+M_GM_N+M_NM_S-3f^2,\\
&~&M_1M_2M_3=M_GM_SM_N-f^2(2M_S+M_N).
\end{eqnarray}
The three physical states can be read as
\begin{equation} \left( \begin{array}{ccc}
|f_0(1710)\rangle\\
|f_0(1500)\rangle\\
|f_0(1370)\rangle
\end{array}\right)
=U\left(\begin{array}{ccc}
|G\rangle\\
|S\rangle\\
|N\rangle
\end{array}\right)
=\left(\begin{array}{ccc}
x_1 & y_1& z_1\\
x_2& y_2& z_2\\
x_3 & y_3 & z_3
\end{array}\right)
\left(\begin{array}{ccc}
|G\rangle\\
|S\rangle\\
|N\rangle
\end{array}\right),
\end{equation}
where 
\begin{equation}
U=\left(\begin{array}{ccc}
(M_1-M_S)(M_1-M_N)C_1& 
(M_1-M_N)fC_1& \sqrt{2}(M_1-M_S)fC_1\\
(M_2-M_S)(M_2-M_N)C_2&
(M_2-M_N)fC_2& \sqrt{2}(M_2-M_S)fC_2\\
(M_3-M_S)(M_3-M_N)C_3&
(M_3-M_N)fC_3& \sqrt{2}(M_3-M_S)fC_3
\end{array}\right)
\end{equation}
with $C_{i(i=1,~2,~3)}=[(M_i-M_{S})^2(M_i-M_N)^2+
(M_i-M_N)^2f^2+2(M_i-M_S)^2f^2]^{-\frac{1}{2}}$.
    
In the above mixing scheme, for the hadronic decays of the 
$f_0(1370)$, $f_0(1500)$ and $f_0(1710)$, neglecting the possible glueball 
component in the final states, we consider the following coupling 
modes as shown in Fig. 1: the coupling of the quarkonia 
components of the three states
to the quarkonia components of the final state 
pseudoscalar mesons, and the coupling of the 
glueball components of the
three states to the quarkonia components of the final state pseudoscalar 
mesons. Performing an elementary $SU(3)$ calculation 
\cite{Tuan,Gao,JRosner,JSch,Haber,H1}, one can get the following equations:

\begin{eqnarray}
&~&\frac{\Gamma(f_0(1500)\rightarrow\eta\eta^\prime)}
{\Gamma(f_0(1500)\rightarrow
\eta\eta)}=\frac{P_{\eta\eta^\prime}}{2P_{\eta\eta}}~
\frac{[4\alpha\beta(\frac{z_2}{\sqrt{2}}-y_2)]^2}
{[(\sqrt{2}\alpha^2z_2+2\beta^2y_2)^2+
2r\cos\theta(\sqrt{2}\alpha^2z_2+2\beta^2y_2) 
x_2+r^2x^2_2)]},\\ 
&~&\frac{\Gamma(f_0(1500)\rightarrow\pi^0\pi^0)}{\Gamma(f_0(1500)\rightarrow
\eta\eta)}=\frac{P_{\pi^0\pi^0}}{P_{\eta\eta}}~
\frac{[\frac{z^2_2}{2}+\sqrt{2}r\cos\theta z_2x_2+r^2x^2_2]}
{[(\sqrt{2}\alpha^2z_2+2\beta^2y_2)^2+
2r\cos\theta(\sqrt{2}\alpha^2z_2+2\beta^2y_2)x_2+r^2x^2_2]},\\
&~&\frac{\Gamma(f_0(1500)\rightarrow K\bar{K})}{\Gamma(f_0(1500)\rightarrow 
\pi\pi)} =\frac{P_{KK}}{3P_{\pi\pi}}~
\frac{[(\frac{z_2}{\sqrt{2}}+y_2)^2+4r\cos\theta(\frac{z_2}{\sqrt{2}}+y_2)x_2
+4r^2x^2_2]}
{[\frac{z^2_2}{2}+\sqrt{2}r\cos\theta z_2x_2+r^2x^2_2]},\\
&~&\frac{\Gamma(f_0(1710)\rightarrow\pi\pi)}{\Gamma(f_0(1710)\rightarrow
 K\bar{K})}=3\frac{P^{\prime}_{\pi\pi}}{P^{\prime}_{KK}}~
\frac{[\frac{z^2_1}{2}+\sqrt{2}r\cos\theta z_1x_1+r^2x^2_1]}
{[(\frac{z_1}{\sqrt{2}}+y_1)^2+
4r\cos\theta(\frac{z_1}{\sqrt{2}}+y_1)+4r^2x^2_1]},\\
&~&\frac{\Gamma(f_0(1710)\rightarrow \eta\eta)}
{\Gamma(f_0(1710)\rightarrow K\bar{K})}=
\frac{P^{\prime}_{\eta\eta}}{P^{\prime}_{KK}}
\frac{[(\sqrt{2}\alpha^2z_1+2\beta^2y_1)^2+
2r\cos\theta(\sqrt{2}\alpha^2z_1+2\beta^2y_1)x_1+r^2x_1]}
{[(\frac{z_1}{\sqrt{2}}+y_1)^2+4r\cos\theta(\frac{z_1}{\sqrt{2}}+y_1)x_1
+4r^2x^2_1]},\\
&~&\frac{\Gamma(f_0(1710)\rightarrow \eta\eta^{\prime})}
{\Gamma(f_0(1710)\rightarrow K\bar{K})}=
\frac{P^{\prime}_{\eta\eta^\prime}}{2P^{\prime}_{KK}}
\frac{[4\alpha\beta(\frac{z_1}{\sqrt{2}}-y_1)]^2}
{[(\frac{z_1}{\sqrt{2}}+y_1)^2+4r\cos\theta(\frac{z_1}{\sqrt{2}}+y_1)x_1
+4r^2x^2_1]},\\
&~&\frac{\Gamma(f_0(1370)\rightarrow \pi\pi)}
{\Gamma(f_0(1370)\rightarrow
K\bar{K})}=3\frac{P^{\prime\prime}_{\pi\pi}}{P^{\prime\prime}_{KK}}
\frac{[\frac{z^2_3}{2}+\sqrt{2}r\cos\theta z_3x_3+r^2x^2_3]} 
{[(\frac{z_3}{\sqrt{2}}+y_3)^2+4r\cos\theta(\frac{z_3}{\sqrt{2}}+y_3)x_3
+4r^2x^2_3]},\\
&~&\frac{\Gamma(f_0(1370)\rightarrow \eta\eta)}
{\Gamma(f_0(1370)\rightarrow K\bar{K})}=
\frac{P^{\prime\prime}_{\eta\eta}}{P^{\prime\prime}_{KK}}
\frac{[(\sqrt{2}\alpha^2z_3+2\beta^2y_3)^2+
2r\cos\theta(\sqrt{2}\alpha^2z_3+2\beta^2y_3)x_3+r^2x^2_3]}
{[(\frac{z_3}{\sqrt{2}}+y_3)^2+4r\cos\theta(\frac{z_3}{\sqrt{2}}+y_3)x_3+4
r^2x^2_3]},
\end{eqnarray}   
where $\alpha=(\cos\theta_p-\sqrt{2}\sin\theta_p)/\sqrt{6}$,
$\beta=(\sin\theta_p+\sqrt{2}\cos\theta_p)/\sqrt{6}$, $\theta_p$ is the 
mixing angle of isoscalar octet-singlet for pseudoscalar mesons; 
$P_{jj^\prime}$ ($P^{\prime}_{jj^\prime}$, $P^{\prime\prime}_{jj^\prime}$)
( $j,j^\prime=\pi,~\eta,~\eta^{\prime},~K$ ) is the momentum of the final 
state meson in the center of mass system for the $jj^\prime$ decays of the 
$f_0(1500)$ ( $f_0(1710)$, $f_0(1370)$ ); $r$ represents the ratio 
of the effective coupling strength of the coupling mode (b) to that of 
the coupling mode (a); $\theta$ is the relative phase between 
the amplitude of the coupling mode (b) and that of the coupling mode (a).

For the two-photon decays of the three states, one can 
get\cite{Parton} 
\begin{eqnarray}
&\Gamma(f_0(1710)\rightarrow \gamma\gamma):\Gamma(f_0(1500)\rightarrow
\gamma\gamma):\Gamma(f_0(1370)\rightarrow \gamma\gamma)=\nonumber\\
&M^3_1(5z_1+\sqrt{2}y_1)^2:M^3_2(5z_2+\sqrt{2}y_2)^2:M^3_3(5z_3+\sqrt{2}y_3)^2
\end{eqnarray}

\section*{III. The results for the cases $M_G>M_S>M_N$ and $M_G>M_N>M_S$  } 
\indent

The decay data relating to the $f_0(1500)$ are as 
follows\cite{Abe96c,Abe98}:

\begin{eqnarray}
&~&\Gamma(f_0(1500)\rightarrow 
\eta\eta^{\prime})/\Gamma(f_0(1500)\rightarrow\eta\eta)=0.84\pm 0.23, 
\nonumber\\ 
&~&\Gamma(f_0(1500)\rightarrow\pi^0\pi^0)/\Gamma(f_0(1500)
\rightarrow\eta\eta)=4.29\pm 0.72 ,
\nonumber\\
&~&\Gamma(f_0(1500)\rightarrow 
K\bar{K})/\Gamma(f_0(1500)\rightarrow\pi\pi)=0.19\pm 0.07.
\end{eqnarray}
 The decay 
datum of the $f_0(1710)$ is \cite{PDG} 
\begin{eqnarray}
 \Gamma(\pi\pi)/\Gamma(K\bar{K})=0.39\pm 0.14.
\end{eqnarray}

Apart from $M_1=1.712$
GeV and $M_2=1.5$ GeV, the central values of the masses of the $f_0(1710)$ 
and $f_0(1500)$, respectively\cite{PDG}, we take the central 
values of the decay data mentioned above and $\theta_p=-19.1^{\circ}$ 
\cite{Coff,Jous} as input. In this way seven parameters, $M_G$, $M_3$, 
$M_N$, $M_S$, $f$, $\theta$ 
and $r$ are unknown. We perform 
numerically to solve the 
unlinear equations $(2)\sim(4),~(7)\sim(10)$ for the cases $M_G > 
M_S > M_N$ and $M_G > M_N > M_S$, respectively. The two solutions 
are presented in Table I. 

Table I shows that in both cases, the
mass of the pure glueball is $1.590$ GeV,
which is in agreement with the lattice QCD simulations which
give $1.55\pm0.05$ GeV \cite{Bali} and $1.63\pm 0.08$ GeV \cite{Lee,Morn}
for the scalar glueball mass. In addition, from Table I the 
masses of the pure $|N\rangle$ and $|S\rangle$ are close to the 
mass of the pure 
glueball in both cases, which implies that a large mixing would exist on 
$f_0(1370)$, 
$f_0(1500)$ and $f_0(1710)$. We note that the mass of $f_0(1370)$ is 
determined to be 
the value of $1.200$ GeV in the case $M_G>M_S>M_N$ 
and $1.380$ GeV in the case $M_G>M_N>M_S$, which is 
also consistent with $1.200\sim1.500$ GeV estimated by PDG\cite{PDG}.

For the 
case $M_G > M_S > M_N$, the numerical form of the unitary matrix $U$ is 
\begin{equation}
U=\left(\begin{array}{ccc}
0.793&0.550&0.263\\
-0.473& 0.830&-0.296\\
0.382&-0.112&-0.917
\end{array}\right).
\end{equation}
The physical states $|f_0(1710)\rangle$, $|f_0(1500)\rangle$ and 
$|f_0(1370)\rangle$ can be read as

\begin{eqnarray}
&~&|f_0(1710)\rangle=0.793|G\rangle+0.550|S\rangle+0.263|N\rangle,\\
&~&|f_0(1500)\rangle=-0.473|G\rangle+0.830|S\rangle-0.296|N\rangle,\\
&~&|f_0(1370)\rangle=0.382|G\rangle-0.112|S\rangle-0.917|N\rangle,
\end{eqnarray}
which indicates that in the case $M_G>M_S>M_N$, $f_0(1710)$ ($f_0(1500)$, 
$f_0(1370)$) contains 
about $63\%$ ($22\%$, $15\%$) glueball component, $30\%$ ($69\%$, 
$1\%$) $s\bar{s}$ component and $7\%$ ($9\%$, $84\%$) 
$(u\bar{u}+d\bar{d})/\sqrt{2}$ component.
In the case $M_G>M_S>M_N$ our results are consistent with the results
given by Ref. \cite{Wein,Stro,Bura}.

Based on Eqs. (7)$\sim$(15) as well as Eqs. (19)$\sim$(21), the numerical 
results relating to the hadronic decays of the $f_0(1710)$, $f_0(1500)$ and 
$f_0(1370)$ in the case $M_G > M_S > M_N$ are shown in the Table II, and 
the two-photon decay width ratio for the three states is given by 
\begin{equation}
\Gamma_{\gamma\gamma}(f_0(1710)):\Gamma_{\gamma\gamma}(f_0(1500)):
\Gamma_{\gamma\gamma}(f_0(1370))=21.917:0.316:38.901.
\end{equation}

For the case $M_G > M_N > M_S$, the numerical form of the unitary matrix
$U$ is
\begin{equation}
U=\left(\begin{array}{ccc}
0.748&0.220&0.626\\
-0.445& -0.527&0.724\\
0.493&-0.816&-0.301
\end{array}\right).
\end{equation}
The physical states $|f_0(1710)\rangle$, $|f_0(1500)\rangle$ and
$|f_0(1370)\rangle$ can be read as  

\begin{eqnarray}
&~&|f_0(1710)\rangle=0.748|G\rangle+0.220|S\rangle+0.626|N\rangle,\\
&~&|f_0(1500)\rangle=-0.445|G\rangle-0.527|S\rangle+0.724|N\rangle,\\
&~&|f_0(1370)\rangle=0.493|G\rangle-0.816|S\rangle-0.301|N\rangle.
\end{eqnarray}
which indicates in the case $M_G>M_N>M_S$, $f_0(1710)$ ($f_0(1500)$,
$f_0(1370)$) contains
about $56\%$ ($20\%$, $24\%$) glueball component, $5\%$ ($28\%$,
$67\%$) $s\bar{s}$ component and $39\%$ ($52\%$, $9\%$)
$(u\bar{u}+d\bar{d})/\sqrt{2}$ component.

Similarly, based on Eqs. (7)$\sim$(15) as well as Eqs. (24)$\sim$(26), the 
numerical results relating to
the hadronic decays of the $f_0(1710)$, $f_0(1500)$ and $f_0(1370)$ are 
shown in the Table III, and the two-photon decay width ratio for the three 
states is given by
\begin{equation}
\Gamma_{\gamma\gamma}(f_0(1710)):\Gamma_{\gamma\gamma}(f_0(1500)):
\Gamma_{\gamma\gamma}(f_0(1370))=59.388:27.908:18.564.
\end{equation}

From Eqs. (18) and (23), in both cases, a large mixing
effect on the $f_0(1710)$, $f_0(1500)$ and $f_0(1370)$ exists, which is 
consistent with our 
above results that in the scalar sector, pure glueball and quarkonia lie 
in a vicinal mass region. The
largest components of $f_0(1370)$ and $f_0(1500)$ are quarkonia and the  
quarkonia content of the $f_0(1500)$ and $f_0(1370)$ differs due to the 
different
mass level order of the bar states $|N\rangle$ and $|S\rangle$, which is 
consistent with the main property of the mass matrix (1) that upon
mixing the higher mass bare state becomes more massive, while the lower
mass bare state becomes less massive (i.e., the mass
splitting between the higher and lower mass bare states
increases as a result of the mixing)\cite{Bura}. The largest 
component of $f_0(1710)$ is glueball, which
supports the argument that $f_J(1710)$ is a mixed $q\bar{q}$ 
glueball  
having a large glueball component if its spin is determined to be  
0\cite{CLOSE}. Furthermore, in both 
cases, the results exhibit destructive interference between the states 
$|N\rangle$ 
and $|S\rangle$ for $f_0(1500)$ while constructive interference between the 
states $|N\rangle$ 
and $|S\rangle$ for $f_0(1710)$ and $f_0(1370)$, which is also 
consistent with the conclusion given by Ref.\cite{CLOSE,Wein,Stro,Bura}.

The predictions about the decays of the three states in two cases 
can provide a stringent consistency check of the two assumptions, 
therefore measurements of the above decay channels of the three states, 
especially the $\eta\eta^\prime$ decay channel
of $f_0(1710)$ and the $\pi\pi$ decay channel of $f_0(1370)$ as well as
the two-photon decays of the three states, 
can be most relevant in clarifying which assumption is really reasonable. 
In addition, it is important to
investigate the nature of the $a_0(1450)$ and $a_0(980)$, since the
isovector scalar state and the isodoublet scalar state $K^{\ast}_0$ can set a
natural mass scale of the ground scalar meson nonet.

\section*{IV. Summary and Conclusions}
\indent

In the scalar glueball-quarkonia mixing framework, we study the 
two-hadronic decays of $f_0(1370)$, $f_0(1500)$ and $f_0(1710)$ 
considering the coupling quarkonia and glueball components of the three 
states to the quarkonia components of the final states pseudoscalar 
mesons. Taking into account two possible assumptions about the mass level 
order of the 
bare states $|G\rangle$, $|S\rangle$ and $|N\rangle$, $M_G>M_S>M_N$ and 
$M_G>M_N>M_S$, we obtain the quarkonia-glueball 
content of the three states by solving the unlinear equations. 

Our conclusions are as follows:

1). In the scalar sector, the pure glueball and quarkonia have comparable 
masses and a significant mixing of glueball with the isoscalar mesons 
exists.

2). The largest component of $f_0(1710)$ is glueball (about $60\%$) 
and the largest component of $f_0(1500)$ ($f_0(1370)$) is quarkonia. 
Which flavor $(u\bar{u}+d\bar{d})/\sqrt{2}$ or $s\bar{s}$ is dominant 
component of $f_0(1500)$ ($f_0(1370)$)
depends on the mass level order of the pure states $|N\rangle$ and 
$|S\rangle$.

3). The interference between $|N\rangle$ and $|S\rangle$ is 
destructive 
for $f_0(1500)$ while constructive for $f_0(1370)$ and $f_0(1710)$.
 
4). The measurements for the $\eta\eta^\prime$ decay channel 
of $f_0(1710)$ and the $\pi\pi$ decay channel of $f_0(1370)$ as well as 
the two-photon decays of the three states would be relevant to judge 
which assumption is really reasonable. Moreover, the confirmation of the 
nature about the $a_0(980)$ and $a_0(1450)$ would be useful to check the 
consistency of two assumptions.

\section*{Acknowledgments}
\indent

We wish to thank Drs. L. Burakovsky and P.R. Page for their useful 
comments on this work. This project is supported by the National Natural 
Science
Foundation of China under Grant No. 19991487, No. 19677205 and Grant No. 
LWTZ-1298 of the Chinese Academy
of Sciences.

\begin{figure}
\centerline{\epsfig{figure=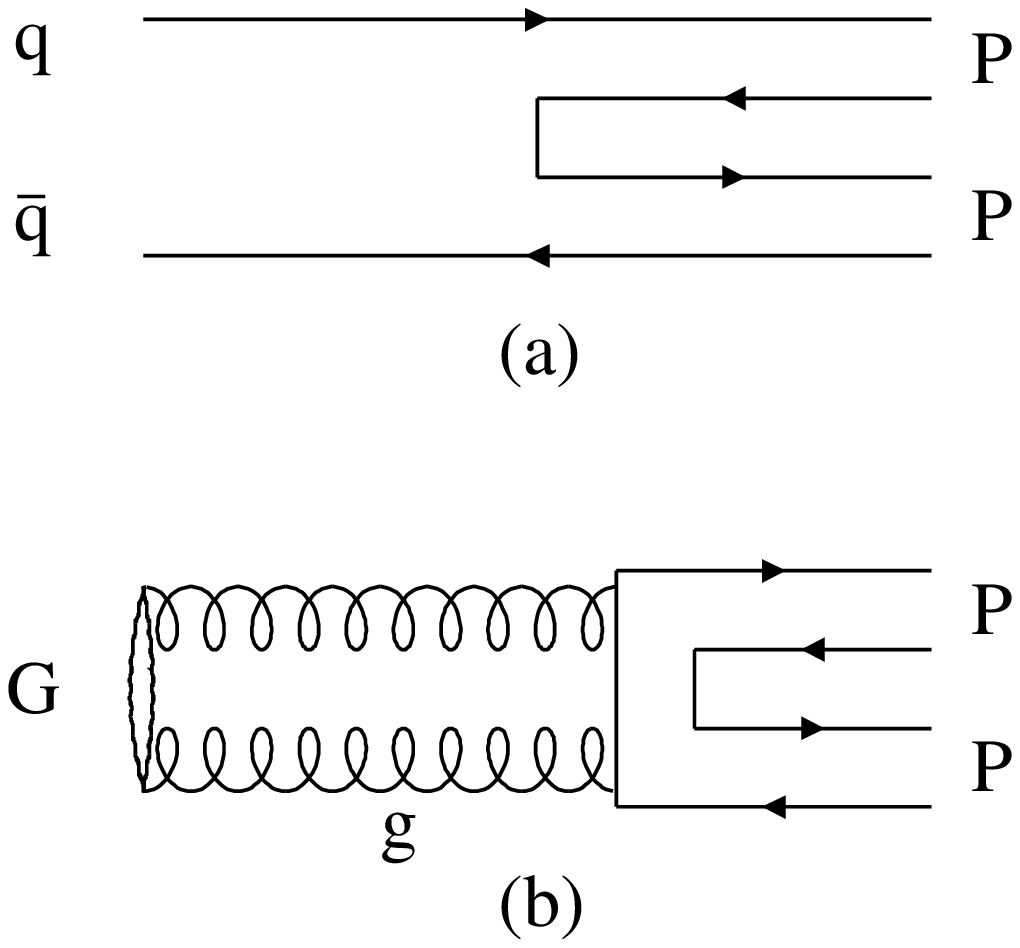}}
\caption{
{\small (a): The coupling of the quarkonia
component
$q\bar{q}$ in the decaying particles to the final state pseudoscalar
mesons $PP$. (b): The coupling of the glueball
component $G$ in the decaying particles to the final state
pseudoscalar mesons $PP$.}}
\end{figure}

\begin{table}

\begin{tabular}{ccc}
Parameters&$M_G > M_S > M_N$&$M_G > M_N > M_S$\\
$M_G$ (GeV) &1.590 &1.590\\
$M_S$ (GeV)&1.560&1.430\\  
$M_N$ (GeV)&1.262&1.572\\  
$M_3$ (GeV)&1.200&1.380\\  
$f$ (GeV)&0.105&0.083\\    
$\cos\theta$ &$0.906$&$-0.974$\\
$r$ &1.220&$0.740$\\
\end{tabular}
\vspace{0.5cm}
\caption{The solutions to the Eqs. $(2)\sim(4),~(7)\sim(10)$ for the
cases $M_G > M_S > M_N$ and $M_G > M_N > M_S$.}
\end{table}  
\vspace{1cm}
\begin{table}

\begin{tabular}{ccc|ccc|ccc}
\multicolumn{3}{c|}{$f_0(1710)$}&
\multicolumn{3}{c|}{$f_0(1500)$}&
\multicolumn{3}{c}{$f_0(1370)$}\\
Modes&Exp.&Theor.&Modes&Exp.&Theor.&Modes&Exp.&Theor.\\
$\frac{\Gamma(\pi\pi)}
{\Gamma(K\bar{K})}$&$0.39\pm0.14$&0.687&$\frac{\Gamma
(\eta\eta^{\prime})}{\Gamma(\eta\eta)}$&$0.84\pm0.23$&0.905&
$\frac{\Gamma(\pi\pi)}
{\Gamma(K\bar{K})}$&&2.858\\
$\frac{\Gamma(\eta\eta)}
{\Gamma(
K\bar{K})}$&&0.216&$\frac{\Gamma(\pi^0\pi^0)}
{\Gamma(\eta\eta)}$&$4.29\pm0.72$&4.225&
$\frac{\Gamma(\eta\eta)}
{\Gamma(K\bar{K})}$&&0.181\\
$\frac{\Gamma( \eta\eta^{\prime})}
{\Gamma(
K\bar{K})}$&&0.005&$\frac{\Gamma(K\bar{K})}
{\Gamma(\pi\pi)}$&$0.19\pm0.07$&0.178&
~&~&\\
\end{tabular}
\vspace{0.5cm}
\caption{The numerical results as well as the experimental data
relating to the decays of the $f_0(1710)$, $f_0(1500)$
and $f_0(1370)$ for the case $M_G>M_S>M_N$.}
\end{table}
\vspace{1cm}
\begin{table}

\begin{tabular}{ccc|ccc|ccc}
\multicolumn{3}{c|}{$f_0(1710)$}&
\multicolumn{3}{c|}{$f_0(1500)$}&
\multicolumn{3}{c}{$f_0(1370)$}\\
Modes&Exp.&Theor.&Modes&Exp.&Theor.&Modes&Exp.&Theor.\\
$\frac{\Gamma(\pi\pi)}{\Gamma(K\bar{K})}$&$0.39\pm0.14$&0.385&$\frac{\Gamma
(\eta\eta^{\prime})}{\Gamma(\eta\eta)}$&$0.84\pm0.23$&0.778&
$\frac{\Gamma(\pi\pi)}
{\Gamma(K\bar{K})}$&&0.454\\
$\frac{\Gamma(\eta\eta)}
{\Gamma(
K\bar{K})}$&&0.181&$\frac{\Gamma(\pi^0\pi^0)}
{\Gamma(\eta\eta)}$&$4.29\pm0.72$&4.269&
$\frac{\Gamma(\eta\eta)}
{\Gamma(K\bar{K})}$&&0.173\\
$\frac{\Gamma(\eta\eta^{\prime})} 
{\Gamma(
K\bar{K})}$&&0.054&$\frac{\Gamma(K\bar{K})}
{\Gamma(\pi\pi)}$&$0.19\pm0.07$&0.151&
~&~&\\
\end{tabular}
\vspace{0.5cm}
\caption{The numerical results as well as the experimental data
relating to the decays of the $f_0(1710)$, $f_0(1500)$
and $f_0(1370)$ for the case $M_G>M_N>M_S$.}
\end{table}

\end{document}